# Effect of recrystallization on diffusion in ultrafine grained Ni

Daria Prokoshkina[a], Leonid Klinger[b], Anna Moros[a], Gerhard Wilde[a],
Eugen Rabkin[b], and Sergiy V. Divinski[a,*]

[a]*Institute of Materials Physics, University of Münster, Wilhelm-Klemm-Str. 10, 48149 Münster, Germany*
[b]*Department of Materials Science and Engineering, Technion–Israel Institute of Technology, 32000 Haifa, Israel*

[*] Corresponding author:     divin@uni-muenster.de

**Abstract**
We studied the effect of recrystallization and grain growth on grain boundary self-diffusion in ultrafine grained Ni prepared by high pressure torsion. Two Ni materials of low (99.6 wt. %) and high (99.99 wt. %) purity levels were compared and the kinetic properties of stationary and mobile grain boundaries were examined in detail. Unusual concentration profiles with characteristic "kinks" were measured for Ni self-diffusion in 99.99 wt. % pure Ni, which is undergoing recrystallization and grain growth during diffusion annealing treatment. This behavior is found to be related to specific kinetic properties of grain boundaries which encompass the recrystallized areas and consume numerous defects of the ultrafine grained matrix during their motion. We proposed a model of self-diffusion in recrystallizing material which takes into account a hierarchy of stationary and moving fast diffusion paths, and satisfactory explains the observed diffusion behavior.

**Keywords**: nickel; severe plastic deformation (SPD); diffusion; recrystallization; grain boundaries (GBs)

## 1. Introduction

It is well known that bulk nano- und ultrafine grained materials exhibit many attractive properties which are superior to those of their coarse grained counterparts. For example, in some cases the refinement of the grain structure increases the yield strength of the materials without significant loss of ductility, which could be promising for technological applications [1]. Although such materials reveal superior mechanical properties, the stability of their microstructure was always an issue [2]. These materials contain a large amount of defects produced by the processing technique (for example, severe plastic deformation), which could induce recovery and recrystallization at comparatively low temperatures.

The kinetics of atomic movement in the grain boundaries (GBs) of bulk nano- and ultrafine grained materials plays an important role in determining their microstructural stability and mechanical properties. In a number of atomistic computer simulation studies it was shown that dislocation plasticity in these materials is controlled by dislocations emission/absorption at the GBs, with both processes being accompanied by diffusion-like atom re-arrangements in the GB region [3]. Moreover, with decreasing grain size the importance of dislocation-mediated processes in plasticity decreases, while the contribution of GB sliding, GB migration, grain rotations, and GB diffusion-controlled grain shape accommodation processes to the total plastic strain increases [4]. The same processes related to the GB self-diffusion are also active during grain growth. Thus, knowing parameters of GB self-diffusion in nanocrystalline materials is imperative for understanding their microstructural stability and mechanical behavior.

Diffusion in evolving microstructures represents an intricate issue, since the correct treatment of experimentally obtained penetration profiles critically depends *inter alia* on properties of stationary and moving GBs. Several theoretical models were proposed to describe diffusion in materials with an evolving microstructure [5-8], which postulated that moving and stationary

GBs exhibit similar diffusivities. An unequivocal proof of this hypothesis is missing so far. It was found that stationary and moving interfaces indeed reveal similar diffusion properties in well-annealed coarse-grained polycrystals [9,10]. Yet the interfaces sweeping through a highly non-equilibrium, deformed material can consume point defects and dislocations, thus temporarily increasing their free volume, energy [11] and diffusivity.

In the present work, ultrafine grained (UFG) Ni of two purity levels has been chosen as a model material for studies of GB self-diffusion. The impurities stabilize the microstructure in the low purity Ni in a wide window of annealing temperatures and times, while the high purity Ni can undergo recrystallization and grain growth under the same annealing conditions. A comparative study of these two materials will establish the effect of interface motion on atomic self-diffusion. Additionally, the properties of moving and stationary interfaces will be compared. The key point is that the kinetic properties of GBs in low- and high-purity Ni in the ultrafine grained state are similar, since the grain refinement in a material with a low impurity content can be considered as a GB "purification" [12]. This idea was recently confirmed for UFG Ni deformed via equal channel angular pressing [13]. Thus, the approach proposed in the present work allows a direct comparison of diffusion properties of moving (recrystallizing high-purity Ni) and stationary (low-purity Ni with stable microstructure) interfaces.

## 2. Experimental procedures
### 2.1. Material
Nickel of 99.99 wt.% (4N) and 99.6 wt.% (2N6) purity levels was employed in this study. The chemical analysis was performed at National Research Technological University "MISiS", Moscow, Russia and the results are presented in Table 1. The average grain size in the initial state was about 20 and 50 μm in Ni of 4N and 2N6 purity, respectively.

Table 1 - Impurity concentrations (in wt. ppm) in the Ni materials under investigation.

| Purity | Co | Cr | Cu | Fe | Mn | Mg | Ti | Zr | Mo | P | S | Si |
|---|---|---|---|---|---|---|---|---|---|---|---|---|
| 4N | 3 | 0.1 | 0.4 | 5 | <0.01 | <0.01 | 0.1 | <0.03 | <0.03 | <20 | 0.6 | 0.4 |
| 2N6 | 60 | 300 | 400 | 800 | 3000 | 300 | 140 | 20 | 2 | <20 | 15 | 250 |

### 2.2. Sample preparation
All samples were cut by spark erosion as 1 mm thick discs and grinded on 1200 SiC paper. The samples were subjected to high-pressure torsion (HPT) at room temperature under the pressure of 2 GPa. Five revolutions with the rate of 0.4 rotations per minute were performed. The ideal equivalent (von Mises) strain, $\varepsilon_{eff}$, was estimated using the following equation,

$$\varepsilon_{eff} = N \frac{2}{\sqrt{3}} \frac{\pi r}{h}, \qquad (1)$$

where $r$, $h$ and $N$ are the sample radius, the final thickness of the sample, and the number of revolutions, respectively. The estimate yielded $\varepsilon_{eff} = 53$ at the half radius of the disc which is considered as a representative value for the subsequent diffusion and calorimetric measurements and microstructure characterization.

Both sides of the HPT-deformed samples were grinded on 1200 SiC paper and cleaned with ethanol in an ultrasonic bath. For diffusion annealing one side of the sample was polished to a mirror-like quality. For all annealing treatments, the samples were wrapped in a nickel foil of 100 μm thickness and 5N-purity and sealed in silica tubes under a purified argon atmosphere. The temperatures were measured and controlled with a certified Ni-NiCr thermocouple with an accuracy of ±1 K.

### 2.3. Hardness measurements
Microhardness was measured on an IndentaMet 1100 Series MicroIndentation Hardness Tester (Buehler) using Vickers' method. A load of 1 kg and a dwell time of 10 s were applied. The hardness was measured along two mutually orthogonal diameters on one side of the HPT disc with a constant step of 0.5 mm. For each sample at least 20 measurements were performed. Nearly constant microhardness values were determined along the diameters, except for the central area and the edges of the sample. These values were excluded from the subsequent averaging.

*2.4. Microstructure characterization*
The microstructure was characterized by a dual-beam Zeiss Gemini focused ion beam (FIB) instrument, operating with a 30 keV liquid-Ga+ source. A beam (Ga+) current of 10 or 50 pA was used providing a channeling contrast on well-polished samples which is sensitive to surface crystallography. In-plane and cross-sectional areas were inspected in detail and a relatively homogeneous microstructure (excluding the center and edge areas) was found.

*2.5. Calorimetric measurements*
Isothermal measurements of the heat release were performed with a Thermal Activity Monitor (TAM) III multi-channel isothermal microcalorimeter (TA Instruments). The sample was first equilibrated slightly below the measurement temperature for about 15 min and then shifted to the measurement position. After about 30 to 45 min all transient processes were accomplished and the heat release was then recorded.

*2.6. Radiotracer diffusion measurements*
The grain boundary Ni self-diffusion was measured using the $^{63}$Ni radiotracer (half-life 100 years) which was available as a 0.5 M hydrochloric solution. The original radiotracer solution was dissolved in distilled water to approach the specific radioactivity of about 4 kBq/μl. Five microliters were dropped on the polished and slightly etched sample surface, dried, then the sample was wrapped in Ni foil and sealed in a silica tube for subsequent diffusion annealing. After diffusion annealing, the samples were reduced in diameter by grinding (by about 1 mm) in order to eliminate effects of lateral surface diffusion. The diffusion penetration profiles were determined by the serial-sectioning technique using a precision parallel grinder with abrasive Mylar foils of 3 μm particle size. The section thickness was determined from the density and the mass reduction by weighing the samples on a microbalance. The section radioactivity (which is proportional to the tracer concentration) was measured by a Liquid scintillation analyzer (TRI-CARB 2910 TR PACKARD, Canberra Co). Other experimental details are similar to those in Ref. [12].

**3. Results**
Typical microstructures of Ni samples of 4N and 2N6 purity after HPT as revealed by FIB are shown in Fig. 1. Similar microstructures were found by inspection of other areas excluding near center and edge regions. The channeling contrast reveals that the grains are almost equiaxed with a slightly prolate shape. The average grain size was determined to be 200±100 nm in 4N Ni, and 100±50 nm in 2N6 one.
The homogeneity of the structure inside the sample was checked by cross-section analysis. The corresponding images showed a similar microstructure throughout the cross section of HPT disks and the grain size was almost the same as determined by in-plane analysis on the disk surface. No explicit shear bands and other deformation defects could be resolved by FIB.

*3.1. Kinetics of recrystallization of 4N Ni*
*3.1.1. Choice of the annealing regimes*
It is well known that the recrystallization processes in pure nanostructured Ni proceed quite fast [13] at low homologous temperatures. In order to find a suitable temperature for kinetic studies in which the microstructure evolves in a reasonable time interval accessible for subsequent diffusion studies, the microhardness of Ni samples was measured after isochronal (17 h) annealing treatments at different temperatures (Fig. 2a). The microhardness of HPT-deformed 2N6 Ni is significantly higher than that of 4N Ni and it remains almost stable in the temperature interval studied.

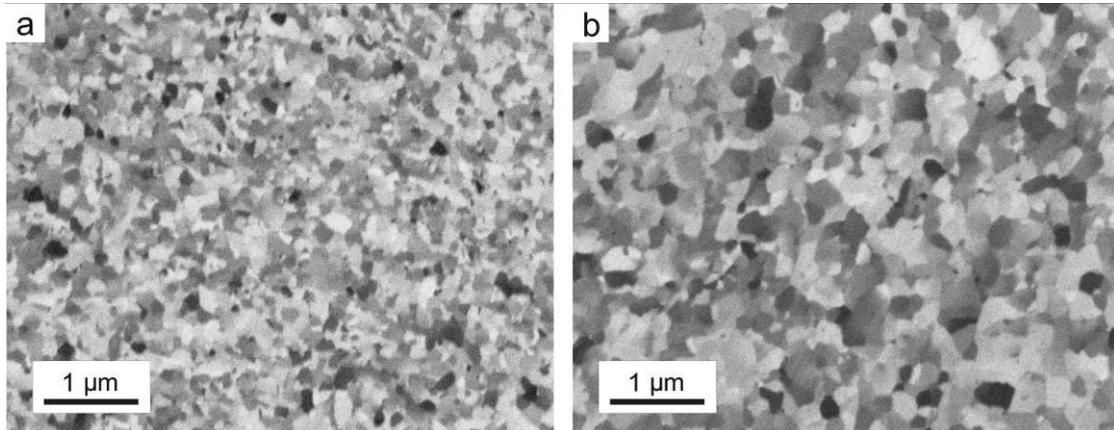

**Figure 1**. UFG microstructure of 2N6 (a) and 4N (b) Ni produced by HPT as revealed via channeling contrast in FIB. These images were obtained at a half radius of the HPT-processed discs.

Whereas a pronounced drop of microhardness of 4N Ni is observed at 423 K, Fig. 2a, the microhardness of 2N6 Ni exhibits subtle changes after all annealing treatments. For more detailed investigations of the recrystallization processes in 4N Ni, the temperature of 423 K was chosen, and the microhardness as function of the annealing time was determined (see Fig. 2b). The microhardness of 4N Ni decreases remarkably after annealing for 20 h and approaches the microhardness of annealed coarse-grained Ni after longer anneals.

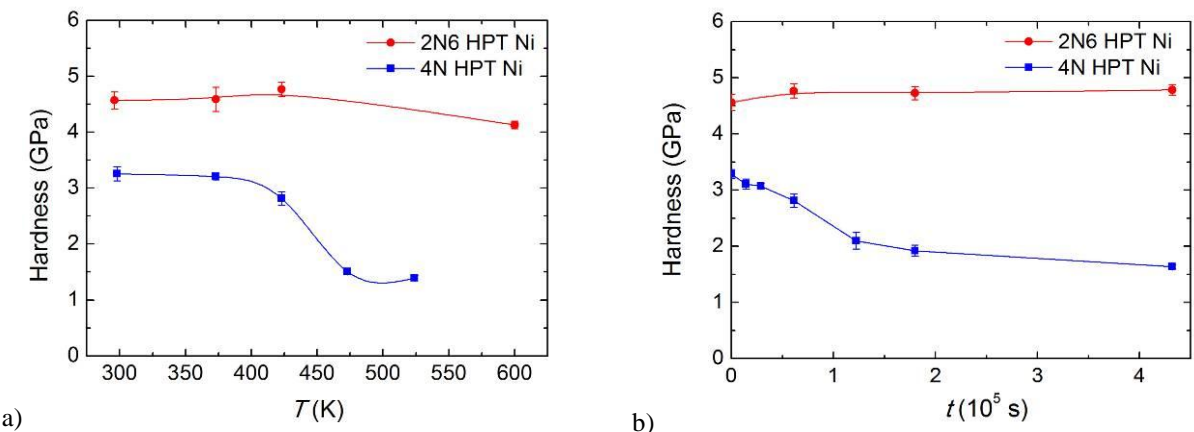

**Figure 2**. Microhardness of 4N (circles) and 2N6-pure (squares) Ni samples as function of the annealing temperature (a) and annealing time at 423 K (b). The lines are drawn as guide for the eye.

In parallel, a slight, but consistent increase of microhardness is observed in 2N6 Ni (about 4 % after 120 h annealing, see Fig. 2b), suggesting that the UFG microstructure remains stable, which is supported by the microstructure observations (see below), with some indications of recovery processes. The microhardness increase of 2N6 Ni at 423 K agrees with the data in Fig. 2a, where a similar increase of the microhardness with increasing temperature is seen. Though this increase is small, it exceeds the experimental error of the hardness measurements. Moreover, the same tendencies were observed in 2N6 nickel deformed via equal channel angular pressing (ECAP) [14] and by cold rolling [15]. Such an increase of the hardness is absent in 4N Ni, but it could be masked by the low temperature recovery and recrystallization processes in this material.

We suggest that this effect is caused by pinning of the dislocations by vacancy-impurity complexes and / or by vacancy clusters. During annealing, a redistribution of impurity atoms (or vacancy-impurity pairs) takes place and these defects accumulate at the dislocations and additionally pin them. The higher the amount of impurities the stronger is the effect.

*3.1.2. Microstructure evolution*

After the hardness measurements, the backside of each sample that was free of indenter marks was polished to a mirror-like quality and the corresponding microstructure was analyzed. At least ten FIB images were recorded for each state. Typical examples of the microstructure evolution in 4N Ni are presented in Fig. 3. No detectable changes of the initial UFG microstructure were established by similar examinations of the 2N6 Ni samples.

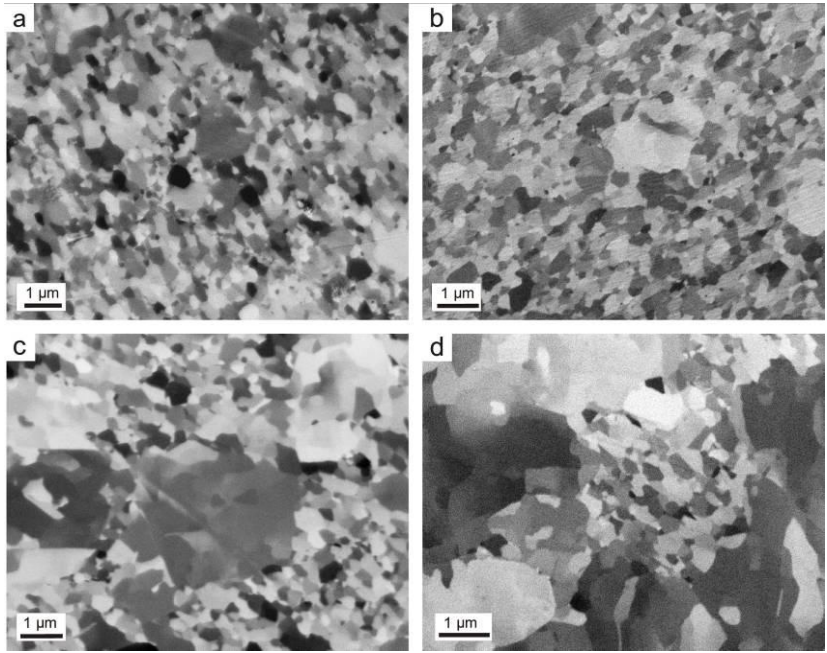

**Figure 3**. FIB images of 4N Ni samples after annealing at 423 K for 8 (a), 17 (b), 34 (c), and 120 (d) hours.

Recrystallization and grain growth at 423 K are explicitly seen in 4N Ni (Fig. 3). However, practically no changes are observed in the residual UFG component of the microstructure (at the given detection limit). The volume fraction, $\lambda$, of the UFG component continuously decreases with increasing annealing time. Cross-section examinations of the annealed HPT sample revealed similar microstructures, which allowed processing the corresponding images and establishing the volume fraction of the recrystallized microstructure component as function of time, Table 2.

Table 2 - The volume fraction of the recrystallized area, $1-\lambda$, as a function of time $t$ after annealing at 423 K

| $t$ (s) | 28800 | 61200 | 122400 | 432000 |
|---|---|---|---|---|
| $1-\lambda$ | 0.194±0.018 | 0.314±0.070 | 0.570±0.059 | 0.872±0.029 |

The recrystallization kinetics could be analyzed in the framework of the Johnson-Mehl-Avrami-Kolmogorov (JMAK) approach [16], which resulted in the following expression, $\lambda(t) = \exp(-B \cdot t^n) = \exp(-3.62 \cdot 10^{-5} \cdot t^{0.85})$. Here the time $t$ is given in seconds. It is seen that the microstructure analysis in terms of the JMAK theory resulted in the exponent $n$ less than unity that may be related to a relatively large uncertainty of the determination of the remaining UFG fraction in FIB images after annealing treatments. Such expression represents the microstructure evolution only approximately, though it is important for the subsequent theoretical treatment of the diffusion problem, see below. The analysis of the recrystallization kinetics in terms of the JMAK theory was further performed via calorimetric measurements which provide volume-averaged data.

*3.1.3. Recovery of defects*
The diffusion kinetics could be affected not only by the microstructure evolution, but by the concomitant recovery processes, too. The heat release during annealing of materials usually corresponds to several consecutive steps, namely, point defect recovery, dislocation annihilation / re-arrangement, formation of sub-grains, and grain growth.

In order to address the defect recovery during annealing treatments, calorimetric measurements were performed. The base-line corrected heat release curves measured for HPT Ni of both purity levels at 390 K during 3 days are presented in Fig. 4a. The obtained heat release curves can only be reasonably fitted by using two exponential functions, shown by the red curves in Fig. 4a. This fact supports the occurrence of two separate processes during heat treatment at 390 K with significantly different (by an order of magnitude) time constants. Moreover, the second (slow) contribution is stronger in 4N Ni, Fig. 4a. It is important that the microstructure remains basically unchanged during annealing at 390 K in Ni samples of both purity levels. Thus, the signal corresponds to the recovery of deformation-induced defects.

The total enthalpy corresponding to the first (fast) process of heat release was calculated by integrating the obtained contributions. Within the accuracy limits, similar values of 0.247 J/g and 0.254 J/g for 4N and 2N6 Ni samples, respectively, were obtained. This process is identified with the annihilation of deformation induced vacancies [17,13]. Taking the vacancy formation enthalpy of 154.3 kJ/mol [18], a vacancy concentration of about $8 \cdot 10^{-5}$ is estimated for both Ni materials. This value agrees well with the literature data for SPD Ni. For example, the vacancy concentration in HPT 4N Ni was reported as $0.2 \cdot 10^{-4}$ to $1.2 \cdot 10^{-4}$, depending on the shear strain [17].

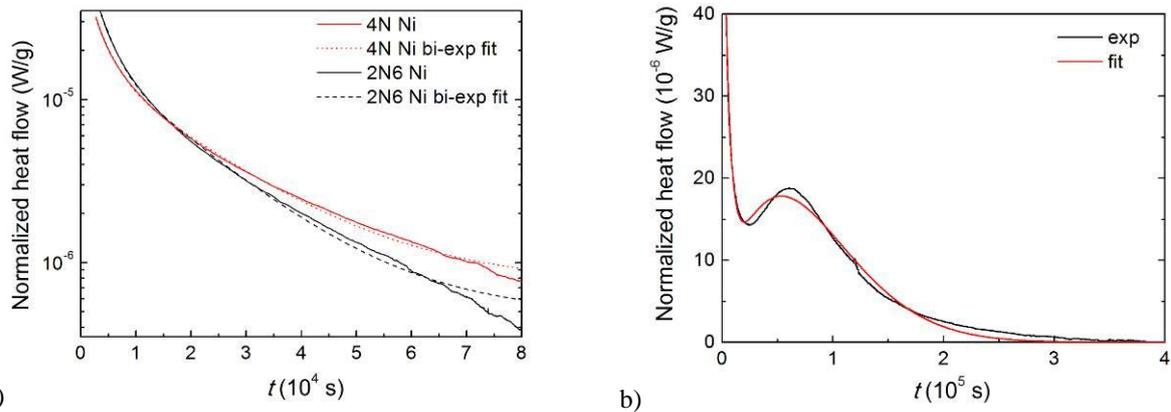

**Figure 4**. Heat release during isothermal annealing of 4N (red curves) and 2N6 (black curves) Ni at 390 K, and their bi-exponential fits (dashed curves) (a), and heat release during isothermal annealing of 4N Ni at 423 K (b). The red curve in (b) represents a fit which accounts for initial exponential decay, and the signal stemming from recrystallization, which is described by the JMAK equation.

Annealing of 4N Ni at 423 K induces recrystallization and its kinetics was examined by recording the corresponding heat release during two weeks, Fig. 4b. According to the microstructure data, recrystallization after such an annealing has to be almost finished. The residual fraction of unrecrystallized UFG matrix should be about 0.4 %. The obtained heat flow-time curve (Fig. 4b) showed a characteristic maximum at times of about 60000 s. A fit of the experimentally measured signal accounting for annihilation of point defects and dislocation rearrangements (i.e. using an exponential decay function) and recrystallization in terms of the JMAK theory is also plotted in Fig. 4b. The best fit is obtained applying the following time dependence of the remaining UFG fraction $\lambda$,

$$\lambda = \exp\left(-6.11\cdot 10^{-9}\cdot t^{1.65}\right). \qquad (2)$$

Here time $t$ is again given in seconds. Due to the local nature of microstructure analyses employed for the derivation of the microstructure-based JMAK equation and a presumably lower accuracy of the corresponding results described in previous section, we will use Eq. (2) for the subsequent theoretical treatment as being more appropriate. Note that no distinct maxima in the heat release curve, Fig. 4b, should be observed if $n$ is less than unity. Although the two functional dependences $\lambda(t)$ determined via the microstructure analysis and calorimetric measurements are obviously different, the absolute difference between corresponding $\lambda$-values is relatively small for the time scale of several days. The main difference arises at long times − no calorimetric signal is measured after holding the sample at 423 K for one week, indicating accomplishment of the recrystallization process, although a low but distinctly observable UFG fraction is still present in the microstructure.

### *3.2. Grain boundary self-diffusion in UFG nickel of 2N6- and 4N-purities*

All radiotracer experiments with HPT Ni samples of 4N and 2N6 purity were carried out at a temperature of 423 K for annealing times of 4, 17 and 120 hours. For these temperature and annealing times the average diffusion distance in the bulk is negligible, and GB diffusion should proceed in the C-type kinetic regime after Harrison's classification [19]. In this regime, the GB diffusion profiles can be linearized in *log c* vs. $y^2$ coordinates, where *c* and *y* are the layer concentration of the radioisotope and the penetration depth, respectively. The penetration profiles measured for 4N Ni are presented in Fig. 5a. These profiles could not be linearized in the *log c* vs. $y^2$ coordinates; this is why we selected a presentation in the coordinates of *log c* vs. *y*. Most measured penetration profiles exhibit characteristic "kinks" (abrupt changes of slope) at depths of several micrometers. Such kinks cannot be described by any GB diffusion model available in the literature.

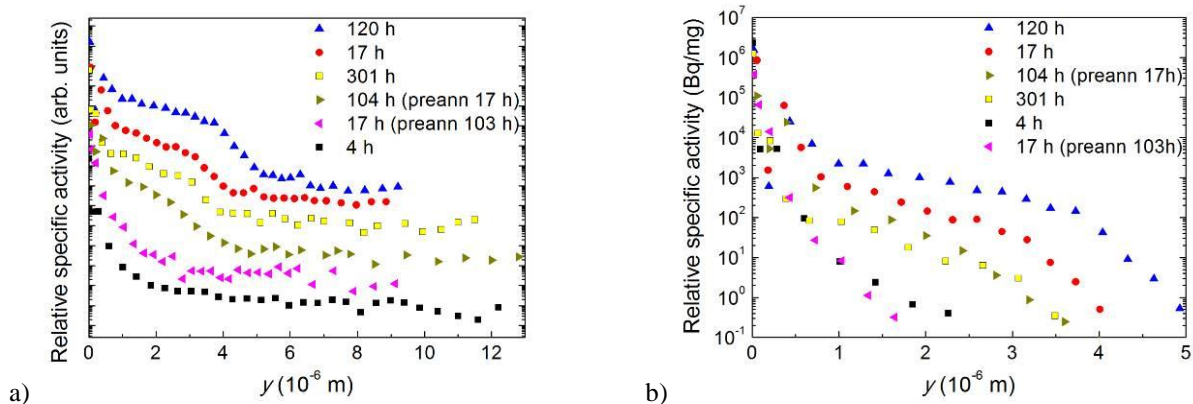

**Figure 5**. Concentration profiles of GB self-diffusion in 4N Ni at 423 K; (a) full set of data; (b) data with the contribution of ultrafast diffusion subtracted.

Fig. 5a suggests that some penetration profiles of self-diffusion in 4N HPT nickel at 423 K reveal unusual "non-Gaussian" behavior with a remarkable kink (especially the profiles measured for 17 and 120 h annealing). This was found to be not the case for less pure 2N6 HPT nickel. The penetration profiles measured in 4N and 2N6 Ni after annealing at 423 K for 120 h are compared in Fig. 6. For reference, the so-called "zero-profiles", i.e. penetration profiles measured just after tracer application without any diffusion treatment, are also given in Fig. 6 for both materials (open symbols), which substantiates the basic difference in diffusion behavior of the two materials – a two-stage penetration is obvious in 4N Ni (with low- and large-depth branches marked I and II in Fig. 6) whereas only one such stage (corresponding to the branch II) is seen in 2N6 Ni.

The branch II is related to ultra-fast diffusion in the UFG matrix akin measured previously in ECAP-processed Cu-Zr [20], Cu [7,21] or Ni [13]. The UFG matrix is stable in 2N6 Ni after annealing at 423 K and a small fraction of it was retained in 4N Ni even after long annealing, see e.g. Fig. 3d. The nature and kinetic properties of the corresponding interfaces will be analyzed elsewhere [22].

In the present work we will concentrate on the first stage, the branch I. Correspondingly, the second stage is simply approximated by a complimentary error-function solution and the pertinent contributions are subtracted from the experimentally determined concentration profiles. The resulting concentration profiles are re-plotted in Fig. 5b.

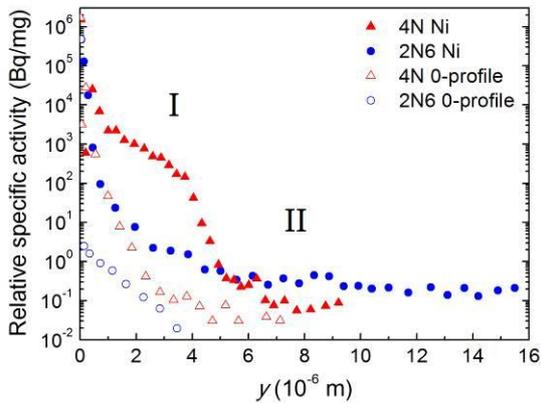

**Figure 6**. Penetration profiles obtained in Ni of 4N and 2N6 purity after annealing at 423 K for 120 h. The so-called zero-profiles, i.e. profiles measured on the corresponding materials without heat treatment, substantiate that the first several points on the concentration profiles (corresponding to depths of about one micrometer) have to be omitted from the subsequent analysis due to contaminating "grinding-in" effects.

## 4. Discussion

The penetration profiles measured in recrystallizing 4N Ni, Fig. 5b, are intricate and to our best knowledge were never reported in the literature so far. In fact, the characteristic kink that appears after diffusion for 17 h is largest for a diffusion annealing duration of 120 h, and again is hardly seen on the penetration profiles measured after annealing for 301 h.

In order to elucidate the nature of this kink, several additional measurements were performed. Two Ni samples were pre-annealed for (*i*) 17 and (*ii*) 103 hours, followed by diffusion annealing treatments for 103 and 17 hours, respectively. The total annealing time was approximately 120 hours, thus similar microstructures are expected at the end of the diffusion heat treatments. The determined penetration profiles are plotted in Fig. 5b, too. Whereas a measurable penetration (for the branch I under consideration) was determined in case (*i*), practically no penetration is seen in case (*ii*). Note, that the branch II is almost not affected by such pre-annealing treatments.

The microstructure analysis shows that after 17 hours of annealing nearly 70 % of UFG grains are left, while this fraction is reduced to about 20 % after annealing for 104 hours. The concentration profile does not exhibit any kink in the latter case, whereas a slight kink is detected in the former. We conclude that when recrystallization is close to be completed, it does not influence the shape of the penetration profile and no measurable radiotracer penetration (in terms of branch I) is observed.

As it was mentioned in the Introduction, three basic models of GB diffusion in the C-type kinetic regime in materials with evolving UFG microstructure are available in the literature: the models suggested by Glaeser and Evans [5], by Amouyal and co-workers [7], and by Rabkin and Klinger [8]. The first model is not suitable for the nanocrystalline materials since it considers boundary migration *perpendicularly* to the diffusion direction that effectively limits the penetration depths to the grain size. Since in the present work the grain size is lower than 300 nm and the penetration depths are deeper than 10 micrometers, the model of Evans and Glaeser cannot be employed. Note that unjustified use of that model for diffusion in nanocrystalline materials may result in orders of magnitude errors in the derived diffusion coefficients [8]. The other two models were developed for nanocrystalline materials, but they do not include the effect of forward (i.e. *parallel* to the diffusion direction) movement of interfaces. Moreover, in both models the diffusion properties of stationary and moving interfaces were assumed to be identical. None of these models predicts the appearance of a kink in the penetration profiles. This kink is most explicitly demonstrated by comparing the concentration profiles measured for 2N6 (no grain growth) and 4N (pronounced recrystallization) Ni at 423 K, see Fig. 6.

Below we summarize the salient features of the present diffusion measurements concerning the appearance of the "kink" in the penetration profiles:

1. The kink appears and becomes first more pronounced with increasing diffusion time in the course of recrystallization, Fig. 5b;

2. Pre-annealing suppresses the kink appearance;

3. No kink is observed if the tracer diffuses in the fully recrystallized matrix;

4. No kink is observed when the tracer diffuses in a stable microstructure (2N6 Ni) and, moreover, branch I is absent in this case, Fig. 6.

These features suggest that the GBs in HPT-processed Ni are of two basic types: GBs with relatively slow (branch I) and relatively fast diffusion coefficients (branch II). When the boundaries corresponding to branch I are stable, as in 2N6 Ni, no significant tracer penetration is observed along them, Fig. 6. The latter reproduces the behavior observed previously in Cu-0.17 wt. % Zr alloy [7] and 99.6 wt. % Ni with stable UFG microstructures produced by ECAP processing.

The situation in 4N Ni during annealing is basically different. At the temperature of 423 K, the recrystallization in this material proceeds discontinuously, with newly nucleated recrystallized grains growing and consuming the remaining UFG matrix, see Fig. 7a. Correspondingly, the deformation-induced GBs (e.g. marked by the magenta lines in Fig. 7a) are retained in the remaining UFG matrix. The recrystallization also produces new GBs between the recrystallized grains (marked in red in Fig. 7a). Since, again, no branch I is observed in the penetration profiles in completely recrystallized Ni, these GBs correspond presumably to general (relaxed) high-angle GBs with a negligible contribution to the total diffusion flux in the present conditions. Additionally, a new type of interfaces is formed in the recrystallizing matrix: the recrystallization front (or envelop of the recrystallized areas) which separates the recrystallized areas and the remaining UFG matrix, marked by the blue line in Fig. 7a.

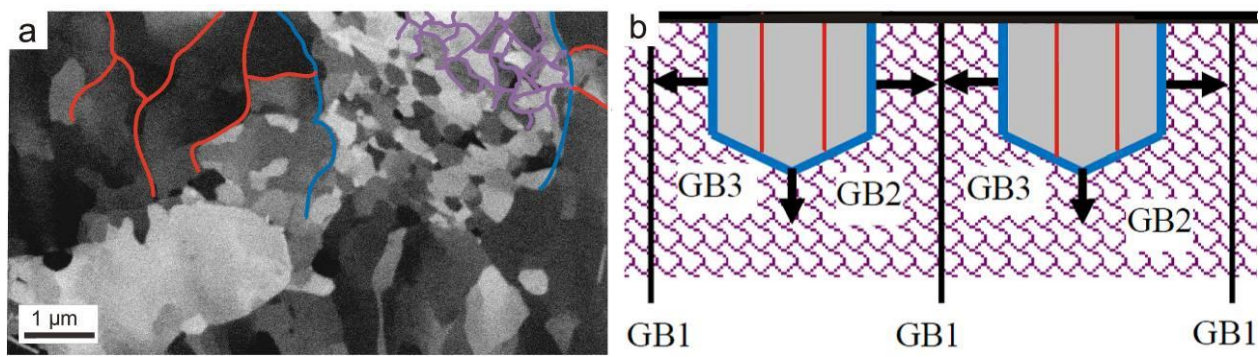

**Figure 7**. Microstructure of 4N Ni after annealing at 423 K for 34 hours (a) and its schematic presentation (b). Three basically different types of (presumably) high-angle GBs are sketched in (a) – the residual deformation-induced boundaries between ultrafine grains (magenta lines), the interfaces between recrystallized grains (red lines), and the recrystallization front (blue line). In (b) the ultrafast GB1 are also shown (black lines) which correspond to the branch II of the penetration profiles.

During recrystallization, the recrystallization front sweeps through the UFG matrix and incorporates the deformation-induced defects, which increases the excess free volume of the interface, and, consequently, its diffusivity. Simultaneously, the defects incorporated at the moving interface may either diffuse towards the interface sinks and annihilate there, or may be left behind in the recrystallized grain in the case the recrystallization front moves too fast.

These assumptions are in a qualitative agreement with the results of both previous studies and the present work. Indeed, the site-specific nanoindentation studies of partly recrystallized ECAP-processed Cu revealed that the hardness of the remaining untransformed UFG matrix is very close to the hardness of as-processed material [23]. This means that little recovery occurs in the UFG matrix during recrystallization, and a high density of deformation-induced defects (except for vacancies) is retained. These defects provide the driving force for continuing recrystallization and, being consumed by the advancing recrystallization front, increase its energy and diffusivity. In the present study, since the migration rate (and, correspondingly, the diffusivity) of the migration front are at maximum at the initial stages of recrystallization, a fast tracer penetration into the sample occurs at this initial stage. Slowing down of the motion of the recrystallization front and a decrease of its diffusivity at the later stages of recrystallization "freeze" the tracer penetration into the sample, which results in the characteristic kink in the penetration profile.

In what follows, we suggest a semi-quantitative model of GB diffusion in recrystallizing material.

### *4.1. Model*

We propose the following model of the GB diffusion during the recrystallization process, see Fig. 7b. As in the experiment, we consider low annealing temperatures when bulk diffusion is frozen. Diffusion occurs along the three kinds of GBs introduced above (Fig. 7b):

- GB1 are the stationary (stable) GBs with high diffusivities $D_1$ and density $1/d_1$, which form a percolating network (their contribution corresponds to the branch II in Fig. 6).
- GB2 are the GBs in the residual UFG matrix with relatively low diffusivities $D_2$ and the density $1/d_2$, which also form a percolating network. Their volume fraction is proportional to the fraction of the residual UFG component, $\lambda$, which decreases with time according to Eq. (2).
- GB3 are the expanding GB loops (recrystallization front) with high diffusivities $D_3$ and density of the loops $1/d_3$, forming a non-percolating network.

The newly-created GBs in the recrystallized areas (red lines in Fig. 7) are considered as general relaxed high-angle ones, whose contribution is negligible in the present conditions.

When the recrystallization front (GB2) moves, the tracer atoms diffusing along GB2 are left behind and remain "frozen" in the bulk. Then, the average layer concentration of radiotracer, $\langle q(y,t) \rangle$ is the sum

$$\langle q(y,t) \rangle = (\delta/d_1)C_1(y,t) + (\lambda\delta/d_2)C_2(y,t) + (\delta C_3(y,t) + q_{\text{bulk}}(y,t))/d_3, \quad (3)$$

where $C_1$, $C_2$ and $C_3$ are the concentrations inside GB1, GB2 and GB3. The parameter $q_{\text{bulk}}$ is the quantity of the diffusing component frozen in the bulk. The diffusion along GB1 and GB2 proceeds independently and the corresponding concentrations $C_1$, $C_2$ are given by the constant source solution of the diffusion problem,

$$C_1(y,t) = C_0 \mathrm{erfc}(y/2\sqrt{D_1 t}), \quad 0 < y < \infty \quad (4)$$

$$C_2(y,t) = C_0 \mathrm{erfc}(y/2\sqrt{D_2 t}), \quad 0 < y < \infty. \quad (5)$$

To simulate the distribution of the radiotracer in GB3 and in the bulk we approximate the shape of the loop (recrystallization front) by a part of a hexagon (Fig. 7b), with vertical side $l_y$ and width $2l_x$ growing with time. The distribution of the component along the expanding loop can be obtained employing the mass conservation condition:

$$\delta \frac{\partial C_3}{\partial t} = \delta D_3 \frac{\partial^2 C_3}{\partial s^2} - V_n(C_3 - C_2\delta/d_2), \qquad 0 < s < l_3(t) \qquad (6)$$

where $s$ is the coordinate along the loop (GB3), $l_3(t)$ is the half of the loop length and $V_n$ is the normal velocity of the GB: for $0<s<l_y$ GB3 are vertical, and for $l_y < s < l_y + l_x 2/\sqrt{3}$ they are inclined to the $y$-axis by $\pi/3$.

Since the radiotracer arrives in the bulk only once it is left behind migrating GB3, its distribution in the bulk is governed by

$$\frac{\partial q_{\text{bulk}}(y,t)}{\partial t} = \begin{cases} V_{n1}(t)C_3(s,t) & 0 < s < l_y \\ 2V_{n2}(t)C_3(s,t) & l_y < s < (l_y + l_x 2/\sqrt{3}) \end{cases} \qquad (7)$$

where $V_{n1}$, $V_{n2}$ are the normal velocities of the vertical and the sloped parts of the crystallization front,

$$V_{n1} = dl_x/dt, \qquad V_{n2} = 0.5 dl_x/dt + 0.5\sqrt{3} dl_y/dt. \qquad (8)$$

We choose the dependences $l_x(t)$ and $l_y(t)$ in a such way that the relative area of the hexagon follows the *JMAK* equation

$$2l_x l_y + l_x^2/\sqrt{3} = (1-\lambda)d_3^2, \quad \text{where } \lambda(t) = \exp(-Bt^n) = \exp(-(t/\tau_0)^n). \qquad (9)$$

The time constant $\tau_0 = B^{-1/n}$ is introduced here just for the sake of convenience, in order to use a dimensionless time in the numerical estimates. The dependencies $l_x$ and $l_y$ on $\lambda$ are shown in Fig. 8.

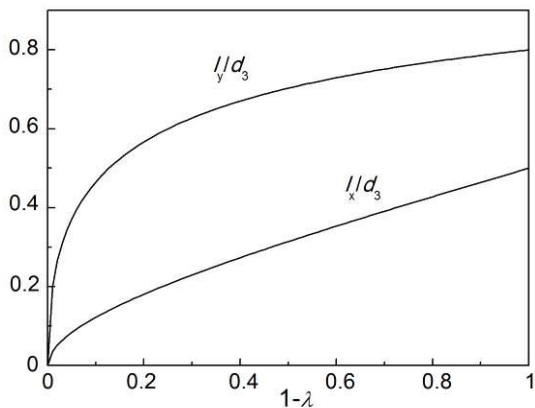

**Figure 8.** Dependence of $l_x$ and $l_y$, normalized to the size of the recrystallized areas, $d_3$, on the fraction of recrystallized matrix, $1-\lambda$.

The analysis of the experimental profiles indicates that the diffusivity of the moving recrystallization front is much higher than that of all other GBs. That is why we also introduce a dependence of the diffusivity of the moving GB3 on its normal velocity. We (somewhat arbitrarily) approximate this dependence by the following formula:

$$D_3(V) = D_2 + (D_{3\max} - D_2)\frac{V^2}{V^2 + V_C^2} \qquad (10)$$

where parameter $V_C$ was chosen to be about $8.3 \cdot 10^{-12}$ m/s. According to Eq. (10), $D_3 \approx D_{3\max} \gg D_2$ at the initial stages of recrystallization (fast recrystallization rate), and $D_3 \approx D_2$ once the recrystallization is close to be completed. The numerical solution of Eqs. (6)–(10) was obtained employing the following values of the relevant parameters: $\tau_0 = 9.5 \cdot 10^4$ s, $n=1.65$ (see Eq.(2)), $D_1 = 1.1 \cdot 10^{-15}$ m$^2$/s, $d_1 = 5 \cdot 10^{-6}$ m, $D_2 = 2.7 \cdot 10^{-21}$ m$^2$/s, $d_2 = 2 \cdot 10^{-7}$ m, $D_{3\max} = 2.4 \cdot 10^{-14}$ m$^2$/s, $d_3 = 5 \cdot 10^{-6}$ m.

It should be noted that all parameters but one here are not arbitrary, but dictated by the *JMAK* equation ($\tau_0$ and $n$), microstructure characteristics ($d_1$, $d_2$, see Fig. 7), the slope of the branch II of the penetration profile ($D_1$), and the GB self-diffusion in equilibrated coarse-grained Ni ($D_2$). Thus, only parameter $D_{3\max}$ was varied to achieve the best agreement between the experimental data and the model prediction. The calculated radiotracer penetration profiles are shown in Fig. 9a in comparison with the experimental profiles for annealing times 4, 17 and 120 h.

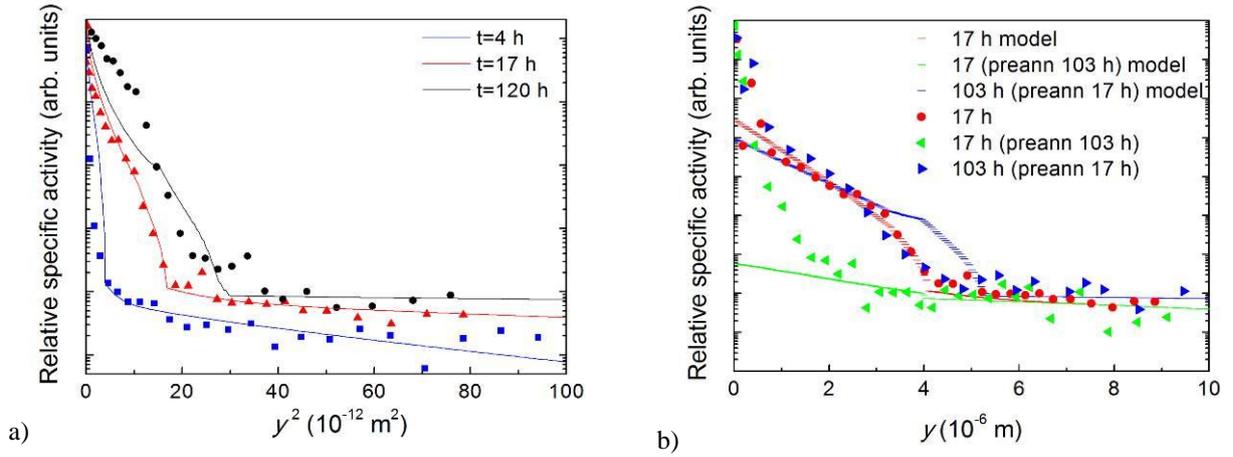

**Figure 9**. Calculated GB self-diffusion penetration profiles in recrystallizing material (curves) and their comparison with experimentally measured profiles (symbols) for GB self-diffusion in HPT-processed Ni without (a) and with the corresponding pre-annealing treatments (b). The pre-annealing times, $t_0$, and the diffusion times, $t$, are indicated. In (b), the experimental profiles measured for $t_0 = 0$h and $t = 17$h (squares), $t_0 = 104$ h and $t = 17$ h (circles), and $t_0 = 17$ h and $t = 103$ h (triangles) are shown for comparison.

Despite its schematic nature, our model correctly reproduces the characteristic kink in branch I of experimentally measured concentration profiles. Moreover, the model also reproduces a significant slowdown of tracer penetration kinetics with increasing annealing time: in both experimentally measured and modeled penetration profiles the increase in annealing time from 17 h to 120 h causes an increase of the average penetration depth of the radiotracer only by a factor of 1.3, well below the factor of 2.7 that would follow from the usual parabolic penetration kinetics. In the framework of our model, the tracer distribution in branch I of the penetration profiles stabilizes after long annealing times exceeding 120 h. For these long times, only branch II of the profile continues to evolve, with its slope decreasing with increasing annealing time. In the experiment, a certain decrease of the amount of radiotracer and of the average penetration depth in branch I was observed after a long annealing time of 301 h (see Fig. 5). This counterintuitive behavior may be related to the outdiffusion through the alternative diffusion paths (such as GB1) of the radiotracer penetrated into the material along the recrystallization front during early stages of recrystallization. This behavior is not reproduced by our model, since it treats the diffusion paths GB1 and GB3 independently. Also, a relatively poor performance of our model for 120 h at shallow penetration depths (up to $3.5 \cdot 10^{-6}$ m) can be clearly seen in Fig. 9a: the slope of the calculated penetration profile is much steeper than observed in the experiment in this range of penetration depths. This discrepancy may be associated with continuing nucleation of new near-surface grains (and, consequently new GB3 diffusion paths) in the course of recrystallization. The new expanding near-surface recrystallyzed grains may increase the effective intermixing rate in the close proximity of the surface. In addition, the recrystallization fronts GB3 may exhibit a range of diffusivities depending on their crystallography, in full analogy with equilibrated GBs in well-annealed bi-crystals [24,25].

The effect of pre-annealing is demonstrated in Fig. 9b. Whereas with the chosen set of input parameters our model reproduces well the profiles measured for a diffusion time $t = 17$ h with and without pre-annealing treatment for $t_0 = 104$ h[1], the experimental profile measured with $t_0 = 17$ h and $t = 103$ h deviates considerably from the model predictions – the penetration depth is even smaller than that for $t_0 = 0$ h and $t = 17$ h, although an opposite trend definitely follows from the model, see Fig. 9b. The reason is a strongly non-monotonous and complex recovery of introduced defects in HPT Ni at 423 K, see Fig. 4b. Annihilation of vacancies occurs very fast during the first several hours of annealing and during this time vacancies significantly contribute to the enhanced diffusion properties of the recrystallization front, GB3. If diffusion starts immediately with annealing (no pre-annealing heat treatment, $t_0 = 0$ h), the diffusing atoms can profit from this enhancement and this is included in the model by a proper choice of the parameters. Otherwise, if $t_0 \neq 0$, vacancies annihilate already after several hours and the diffusion enhancement of the moving recrystallization front is significantly decreased. The last feature is not captured by our model which was formulated as simple as possible introducing only one time constant which is the same for recovery of the diffusion enhancement of the recrystallization front, Eq. (10), and for the recrystallization process itself, Eq. (2). Accounting for the intricate character of the recovery kinetics of deformation-induced defects is relatively straightforward, but would result in a more complicated and less transparent model. Therefore, we decided to limit ourselves by the current formulation of the model, Eqs. (2)–(10), which, though being relatively simple, provides a satisfactory agreement with the experiment.

It is interesting that no characteristic kinks in the penetration profiles similar to those found in this work were observed in previous studies of Ni diffusion in UFG Cu processed by ECAP, in which the recrystallization also occurred simultaneously with diffusion [7]. We believe that this is related to the maximum density of defects accumulated by the material during severe plastic deformation. Indeed, due to the lower melting point of Cu in comparison with Ni, the processing (room) temperature corresponds to a higher homologous temperature in Cu. This means that partial recovery of defects and, to some extent, dynamic recrystallization occur in Cu already during processing. Thus, the final density of defects in Cu processed by severe plastic deformation should be lower than in more refractory Ni. Consequently, the migrating recrystallization front in Cu absorbs fewer defects than in Ni, and the corresponding surge in diffusivity should be also lower. A super-fast diffusivity along the recrystallization front is instrumental for the formation of kinks in the tracer penetration profiles. This is illustrated in Fig. 10, in which several concentration profiles calculated in the framework of the present model with different values of $D_{3max}$ (initial diffusivity along the recrystallization front, GB3) are shown. While a minor kink is still observed for $D_{3max} = 1.1 \cdot 10^{-14}$ m$^2$/s, the penetration profiles become fully monotonous for $D_{3max} < 5.6 \cdot 10^{-15}$ m$^2$/s. It should be emphasized that the latter diffusivities are still faster than the diffusivities of percolating diffusion paths GB1, which were considered as "ultrafast" by Amouyal et al. [7], yet they do not produce any kinks. Thus we conclude that the different diffusion behavior of UFG Cu and Ni is related to a lower relative increase in diffusivity of the advancing recrystallization front in Cu.

The developed model provides further arguments in favour of Eq. (2) over the predictions of the JMAK kinetics based on microstructure analysis. According to the developed model, it is not just the *recrystallized fraction* itself, but the *recrystallization rate* which plays a critical role in accelerating diffusion. In this case the existence of residual unrecrystallized areas, whose fraction is practically constant for long times, does not contribute to short-circuit diffusion in terms of branch I and Eq. (2) is appropriate to quantify the effect of the microstructure evolution on the GB diffusion.

---

[1] The near surface points at the depths below 1 micrometer are affected by the sectioning method and have to be omitted.

Finally, the relative stability of the microstructures in Ni of two different purity levels has to be critically discussed. While recrystallization and grain growth is observed in Ni of 4N purity, the HPT-produced UFG microstructure remains stable in Ni of 2N6 purity. One may argue that it is segregation of impurities in less pure Ni which stabilized the UFG microstructure of the latter. However, similar self-diffusion rates along the interfaces in the (unchanged) UFG matrices in both materials follow from the present data and analysis. This apparent contradiction is resolved in view of the recent data on the "purification effect" of grain refinement [12]. While the given amount of impurities has a large effect on grain boundary self-diffusion for coarse-grained materials, their enrichment at the GBs drastically decreases with decreasing grain size, which results in GB self-diffusion rates in the UFG materials with a low level of purity which are comparable with those in pure material. We suggest that particles of impurity-rich phases effectively pin the GBs and slow down the recrystallization rates in UFG materials of low purity levels. The role of second-phase particles was clearly demonstrated in the case of ECAP-processed Cu-0.17 wt. % Zr alloy [7] and for a nano-scaled Cu-W alloy [26]. Pinning by triple junctions due to their low mobility could also contribute to such a behavior [27].

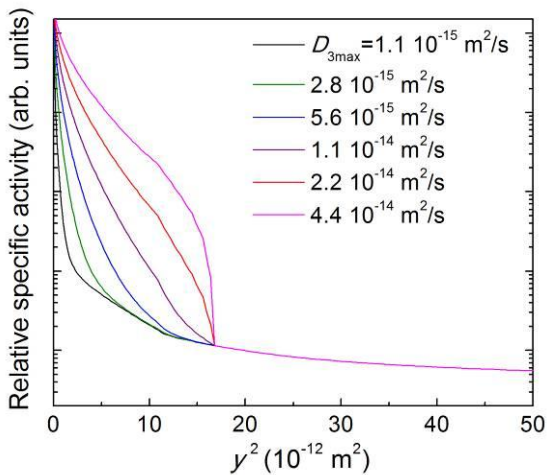

**Figure 10**. Calculated GB self-diffusion penetration profiles in recrystallizing material after annealing for 17 h and for different values of diffusivities along the recrystallization front.

## 5. Summary

The effect of recrystallization and grain growth on self-diffusion was investigated in UFG Ni of two purity levels (99.6 and 99.99 wt. %) prepared by HPT processing. Based on the results of the present study, the following conclusions can be drawn:

1. The microstructure of Ni of both levels of purity produced by HPT is to a large extent homogeneous with almost equiaxed grains. The effect of purity on the UFG structure can be reduced to the difference in grain sizes, which were found to be about 100 and 200 nm in Ni of 2N6 and 4N purity, respectively.

2. The time and temperature range in which the microstructure of 2N6 Ni is stable and recrystallization in 4N Ni occurs was determined by microhardness measurements. The hardness of as-prepared samples of 2N6 Ni was remarkably higher than that of 4N Ni, which is consistent with the difference in the grain size and dislocation pinning by impurity-rich particles in 2N6 Ni.

3. Recrystallization kinetics in 4N Ni during annealing at 423 K was investigated and analyzed in the framework of the JMAK approach. The unreasonably small value of the JMAK exponent obtained from microstructure analyses was corrected using sensitive isothermal microcalorimetry.

4. Using the calorimetric measurements, the recovery of the defects that were induced by HPT was examined during heat treatment. A deformation-induced vacancy concentration of about $8 \cdot 10^{-5}$ is estimated for Ni samples of both purity levels.

5. A comparative radiotracer study of the GB self-diffusion in Ni samples of both purity levels was performed and the kinetic properties of stationary and mobile GBs were explored.

6. Measured concentration profiles of GB self-diffusion in 4N Ni, which was undergoing discontinuous recrystallization and grain growth during diffusion annealing, exhibited a unique shape, which has never been reported previously. The profiles exhibited two distinct regions with steep and low slopes (branch I and II, respectively). Branch II was associated with fast GB diffusion along deformation-modified non-relaxed GBs; such part was also found in 2N6 Ni. Branch I (which is absent in 2N6 Ni) exhibited characteristic "kinks" and was related to specific kinetic properties of the GBs encompassing the recrystallized areas and consuming deformation-induced defects of the UFG matrix during their motion (recrystallization fronts).

7. A semi-quantitative model which satisfactory explains the behavior of the experimentally measured penetration profiles was suggested.

8. The diffusion coefficient of a *moving* recrystallization front is found to be strongly enhanced as compared both with that of general (relaxed) high-angle GBs as they can be found in well-annealed coarse-grained polycrystals, and with that of non-relaxed GBs responsible for the branch II of the penetration profiles. This super-fast diffusivity is a result of the consumption of the deformation-induced defects in the UFG matrix by the moving front.

**Acknowledgment**

This research was supported by a research Grant from GIF, the German-Israeli Foundation for Scientific Research and Development (Grant No. G-1037-38.10/2009) and partially by the Deutsche Forschungsgemeinschaft.